# Risk Management for Complex Calculations:
## EuSpRIG Best Practices in Hybrid Applications

**Deborah Cernauskas**
dcernauskas@i4mt.org

**Andrew Kumiega**
kumiega@comcast.net

**Ben Van Vliet**
bvanvliet@i4mt.org

**Abstract**

As the need for advanced, interactive mathematical models has increased, user/programmers are increasingly choosing the MatLab® scripting language over spreadsheets. However, applications developed in these tools have high error risk, and no best practices exist. We recommend that advanced, highly mathematical applications incorporate these tools with spreadsheets into hybrid applications, where developers can apply EuSpRIG best practices. Development of hybrid applications can reduce the potential for errors, shorten development time, and enable higher level operations. We believe that hybrid applications are the future and over the course of this paper, we apply and extend spreadsheet best practices to reduce or prevent risks in hybrid Excel/MatLab® applications.

## I.  Introduction

The spreadsheet is an enabling technology. Spreadsheets, based on Dan Bricklin's VisiCalc, enable user/programmers to organize, collect, store, analyze, perform calculations and report data. With spreadsheets, users need not be computer scientists to develop sophisticated algorithms in code, freeing them to focus on their business strategies and quantifiable algorithms.

To keep up with the demand for more and more sophisticated analysis, spreadsheet capability has evolved greatly since the VisiCalc and Lotus 1-2-3 days. As the capability of the spreadsheet has grown, so too have the sizes of spreadsheet applications, the complexities of calculations and, therefore, the probability of errors.

Material errors in spreadsheets arise often because user/programmers are not well-versed in proper programming practices, despite Boehm's findings that good architectural practices can reduce the cost of program implementation by avoiding defects (Boehm, 2001). Rafftensperger (2001) estimates that 90% of spreadsheets have errors with consequences ranging from mild to severe. Studies by Davies and Ikin (1987) and Cragg and King (1993) found very informal spreadsheet development processes, most of which persist today.

There are several proven techniques that can reduce errors in and shorten development time of spreadsheets. Many of these are best practices documented within the EuSpRIG sphere. For operations beyond the capability of the spreadsheet, users can incorporate VBA code or third-party XLLs. Best programming and testing practices can apply to the coding of VBA macros and function libraries, as well as XLL libraries.

Creating new VBA functions promotes reusability, but VBA is no panacea. Here are two examples. Consider creating a Black-Scholes option pricing model or q-q plot (a quantile-quantile (q-q) plot is a graphical technique for evaluating whether two sets of data come from populations with the same distribution) in VBA code. A Black-Scholes function requires fifteen to twenty lines of code; a q-q plot substantially more. Although VBA has increased the functionality of the spreadsheet, it may not reduce development time, nor does it necessarily reduce the probability of errors. Today's users demand more.

## II.     MathWorks' MatLab

In more and more industries, the level of quantitative inquiry is passing by the technology that originally enabled it. For example, Excel can only handle matrices up to 52x52 in size. For larger matrix algorithms and greater capability, engineers are increasingly using MathWorks' MatLab software (which incidentally also traces its origins back to VisiCalc) for complex calculations. MatLab is a high-level language and interactive development environment that enables users to perform computationally intensive tasks. These tasks can range from financial modeling, to computational biology, to simulation and optimization, to signal processing and control system design. According to the MathWorks website, MatLab is "used today by more than 500,000 engineers and scientists and by more than 2,000 financial companies worldwide." Interestingly, the website goes on to explain that "professionals rely on MatLab to reduce development time, minimize costs and risks, and integrate new models." The same could be said for spreadsheets. Conspicuous in their absence, however, are any documented best practices or proofs to support MathWorks' claim of risk reduction.

Like the spreadsheet, MatLab is an enabling technology. Like spreadsheets, MatLab applications are also subject to risk of errors and fraud. Like enhancers of spreadsheet capability such asVBA or XLLs, though, MatLab is no panacea. MatLab has risks, including:

- Difficulty in testing
- interim calculations and results due to a complex data interface.
- Lack of top to bottom, left to right structure for calculations.
- Ability to build applications interactively by manipulating arrays through a simple command-line structure. This produces code that can ramble as sample code gets copied and pasted into executable code.
- Lack of audited calculations and workflow due to a lack of data structures and documentation outside of the code.
- Absence of best practice development processes (this is very similar to Excel prior to EuSpRIG).

These risks do not (or no longer should) exist in spreadsheets. Which is to point out the following: if MatLab's capabilities could be accessed through a spreadsheet, then EuSpRIG best practices could control these risks. This is now possible.

Through MatLab's new Excel Link users can integrate MatLab's mathematical and graphical capabilities into an Excel spreadsheet. Excel Link is a giant leap forward in facilitating the use of spreadsheets for complex calculations with efficiency, transparency and error reduction. Excel Link allows spreadsheet users to harness the power of MatLab's extensive list of functions for modeling, statistics, and graphing in spreadsheet cells and VBA modules. Excel Link should be an equally large leap forward for MatLab users by providing their applications with a well-defined and easily auditable data interface. Excel Link allows MatLab users to use Excel for the data storage and presentation and MatLab for complex calculations. For example, implementing a Black-Scholes option pricing model or a q-q plot requires a single function call in MatLab, much simpler and more error-proof than creating VBA code. Excel Link allows spreadsheet capability to grow exponentially. At the same time, accessing MatLab through Excel also allows for control of MatLab risks through application of EuSpRIG best practices.

Once the prototype is complete, the MatLab code can be compiled into a .dll by using MatLab Builder for Excel. The .dll then ensures a high quality application with secure calculation code.

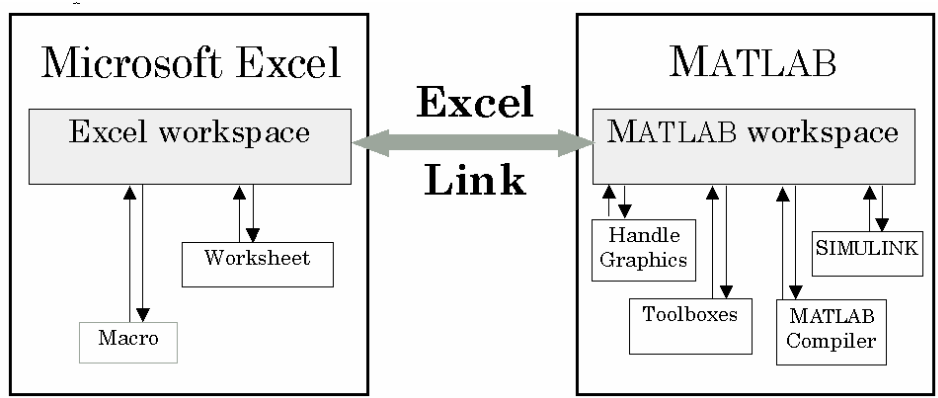

Figure 1: Excel and MatLab (source: www.MathWorks.com)

To generalize, incorporating a commercial-off-the-shelf (COTS) component, such as MatLab, expands the complexity of analytical work that can be done in spreadsheets. Through COTS components, users can implement calculations once beyond the ability of the spreadsheet, while controlling risks inherent in the COTS development environment within a EuSpRIG best practices framework.

### III. Complex Calculations in Banking

CODA, a financial intelligence firm, reports that 95% of U.S. corporations use spreadsheets in financial reporting. For American financial institutions, the enactment of

Sarbanes-Oxley in 2002 in addition to the recommendations of Basel II (Basel II's three pillars promote stability of the international financial system through minimum capital requirements, supervisory review and market discipline) has forced firms to look at how spreadsheets are used in financial reporting systems and to develop methods and procedures to reduce errors and increase security.

The same rules also apply to MatLab applications used in finance, since these applications regularly calculate profit and loss and risk reserves. Fraud or material errors in these spreadsheets can have significant consequences for such firms and their employees, including adverse audit opinions; reductions in stock price; and the possibility of fines and even incarceration.

## IV.    COTS Components:  A Method of Defect Reduction & Prevention

The use of high level mathematics for derivatives pricing and forecasting of market, interest rate and credit risk, is forcing financial engineers to move beyond spreadsheets to more robust, rapid prototyping calculation engines available in COTS software. (COTS applications allow a user to quickly perform complex calculations faster than full development in C++. This is why MatLab is the tool of choice beyond Excel.) Implementing tested, COTS components reduces risk.

Software quality assurance should focus on preventive measures rather than testing. An alternative method to reduce and prevent defect is the use of re-usable software components, which accelerate development and are a big step towards achieving better, faster, cheaper software.

A recent study by Mohagheghi, et al. found that reusable software components have a lower rate of defects given the size of the code than non-reusable components. The same study found that non-reusable components have more errors which are correlated with the amount of the code.

COTS components, like MatLab have several advantages over non-reusable VBA components mainly due to the larger pre-built and tested functions (See Table 2). These include, but are not limited to, fewer defects, lower development costs, and shortened development time.

## V.    MatLab and Excel:  Similarities & Differences

In Excel, shielding users from analytical complexities, is done through user controlled buttons, spinners and macros. Macros can also be coded in MatLab, which allows for more complex calculations often used in financial applications.

| Similarities ||
|---|---|
| **Excel** | **MatLab** |
| No database. Programs use flat files. | Same |
| Interactive | Interactive, but cumbersome |
| No versioning, i.e. most spreadsheets are single use, custom analytical tools. | No versioning, i.e. most MatLab programs are single use, custom analytical tools. |
| Does not have a nice GUI, but one can be built. | Same |
| | |
| **Differences** | |
| Can create a nice audit trail to show intermediate steps in a calculation. | Does not have a nice audit trail. |
| Matrix computations are limited to a size of 52 x 52. | Matrix calculations limited to computer memory size. |
| Regression analysis is easy to use but has limited statistics on equation fit. | Has an extensive list of functions available for assessing the fit of a regression equation. |
| Time series analysis is severely limited due to matrix limitations. | Has an extensive list of time series analysis functions. No limitations due to matrices. |
| Ability to create visual tools and dashboards for analysis and monitoring. | Difficult to program visual tools and dashboards. |

Table 1: Similarities between Excel and MatLab

The capability of the spreadsheet environment is expanded exponentially through Excel Link to MatLab. The MatLab software suite includes a variety of specialized functions in toolboxes such as financial time series, financial functions, and garch modeling.

The amount of re-usable code, i.e. pre-built functions, for math and finance is substantially greater in MatLab than in Excel. This is shown by the number of pre-built functions as listed:

| **Business Use** | **Excel** | **MatLab** |
|---|---|---|
| Optimazation | Solver | Optimization Tool Box |
| Genetic Algorithm | Solver Pro | Genetic Algorithm and Direct Search Tool Box |
| Time Series | No pre-built functions | Signal Processing Tool Box<br>Neural Network Tool Box<br>Model Predictive Tool Box<br>Wavelet Tool Box<br>Garch Tool Box for Forecasting |
| Control Analysis | No pre-built functions | Control System Tool Box<br>Fuzzy Logic Tool Box<br>Robust Control Tool Box |
| Financial Calculations | Basic financial functions | Financial Tool Box<br>Financial Derivative Tool Box<br>Fixed Income Tool Box |

|  |  | Spline Tool Box |
|---|---|---|

Table 2: Comparison of functionality in Excel and MatLab

## VI.     Incorporating MatLab into a Spreadsheet

Employing MatLab commands in the Excel environment is straightforward since MatLab installs as a toolbar into Excel as shown in Figure 2.

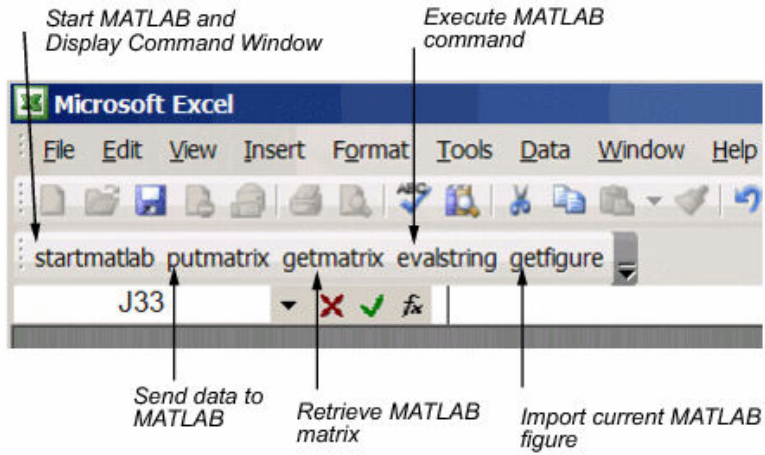

Figure 2: The MatLab toolbar in Excel (source: www.MathWorks.com)

All MatLab functions, commands and programs can be called using the MLEvalString command.  Furthermore, once the MatLab code is completed it can be turned into a VBA module or a .dll file, via MatLab Builder for Excel.   This ensures that the MatLab calculations can be locked and distributed so that end users cannot modify the algorithms.  Figure 3 shows the Excel calls to access MatLab functions.

## Link Management Functions

Excel Link provides four link management functions to initialize, start, and stop Excel Link and MATLAB.

| Function | Purpose |
|---|---|
| matlabinit | Initialize Excel Link and start MATLAB process. |
| MLAutoStart | Automatically start MATLAB process. |
| MLClose | Terminate MATLAB process. |
| MLOpen | Start MATLAB process. |

You can invoke any link management function except matlabinit as a worksheet cell formula or in a macro. You invoke matlabinit from the Excel **Tools Macro** menu or in a macro subroutine.

Use MLAutoStart to toggle automatic startup. If you install and configure Excel Link according to the default instructions, Excel Link and MATLAB automatically start every time you start Excel. If you choose manual startup, use matlabinit to initialize Excel Link and start MATLAB.

Use MLClose to stop MATLAB without stopping Excel, and use MLOpen or matlabinit to restart MATLAB in the same Excel session.

## Data Management Functions

Excel Link provides the following data management functions to copy data between Excel and MATLAB and to execute MATLAB commands from Excel.

| Function | Purpose |
|---|---|
| matlabfcn | Evaluate MATLAB command given Excel data. |
| matlabsub | Evaluate MATLAB command given Excel data and designate output location. |
| MLAppendMatrix | Create or append MATLAB matrix with data from Excel worksheet. |
| MLDeleteMatrix | Delete MATLAB matrix. |
| MLEvalString | Evaluate command in MATLAB. |
| MLGetFigure | Import current MATLAB figure into Excel spreadsheet. |
| MLGetMatrix | Write contents of MATLAB matrix in Excel worksheet. |
| MLGetVar | Write contents of MATLAB matrix in Excel VBA variable. |
| MLPutMatrix | Create or overwrite MATLAB matrix with data from Excel worksheet. |
| MLPutVar | Create or overwrite MATLAB matrix with data from Excel VBA variable. |
| MLShowMatlabErrors | Used by MLEvalString to return standard Excel Link errors or full MATLAB errors. |
| MLStartDir | Specify current working directory of MATLAB after startup. |
| MLUseFullDesktop | Specify whether to use full MATLAB desktop or only Command window. |

You can invoke any data management function except MLPutVar as a worksheet cell formula or in a macro. You can invoke MLPutVar only in a macro.

Figure 3: MatLab Link Management Functions and Data Management Functions

Figure 4 shows two columns of data in an Excel sheet. There are several ways to put this data into MatLab, but the most straightforward way is to use the "matlabsub" function. In cell D6, the following MatLab command is entered:
matlabsub("mean","E6",A4:A1003")

Figure 4: Applying MatLab functions to Excel data

The first argument to the function is the actual MatLab function that will be executed; the second argument is the starting cell location for the function output; and the third argument gives the cell locations of the input data).

Alternatively, the Excel data can be put into a MatLab matrix by using the MLPutMatrix function, which has two arguments: the MatLab variable name and the Excel cell address or range name. In Figure 5, the input data in cells A4:B1003 are assigned to a MatLab matrix called data through the named cell range DATA. The command in cell D15 gives the MatLab name "x" to the data in cells A4:A1003. The command in cell D17 calculates the mean of x and the command in D19 writes out the mean of x to cell E19.

Obviously, the power of MatLab commands within the Excel environment comes from the ability to execute the more complex MatLab commands.

|    | A | B  | C | D | E        | F        | G              | H           | I | J        | K        |
|----|---|----|---|---|----------|----------|----------------|-------------|---|----------|----------|
| 1  |   |    |   |   |          |          |                |             |   |          |          |
| 2  |   |    |   |   |          |          |                |             |   |          |          |
| 3  | X | Y  |   |   |          |          |                |             |   |          |          |
| 4  | 0 | 21 |   |   | 0        |          |                |             |   |          |          |
| 5  | 3 | 18 |   |   |          |          |                |             |   |          |          |
| 6  | 1 | 34 |   |   | 0        | 1.96     | matlabsub("mean","E6",A4:A1003) |   |   |          |          |
| 7  | 0 | 31 |   |   |          |          |                |             |   |          |          |
| 8  | 1 | 29 |   |   | 0        |          | matlabsub("cov","J8",A4:B1003) |    |   | 2.032432 | 0.191071 |
| 9  | 1 | 23 |   |   |          |          |                |             |   | 0.191071 | 22.83675 |
| 10 | 1 | 20 |   |   |          |          |                |             |   |          |          |
| 11 | 3 | 26 |   |   | 0        | 22.83675 | matlabsub("var","E11",B4:B1003) |   |   |          |          |
| 12 | 4 | 19 |   |   |          |          |                |             |   |          |          |
| 13 | 3 | 15 |   |   | 0        |          | MLPutMatrix("data",DATA) |       |   |          |          |
| 14 | 2 | 19 |   |   |          |          |                |             |   |          |          |
| 15 | 0 | 31 |   |   | 0        |          | MLEvalString("x=data(:,1)") |    |   |          |          |
| 16 | 2 | 26 |   |   |          |          |                |             |   |          |          |
| 17 | 0 | 22 |   |   | 0        |          | MLEvalString("m=mean(x)") |      |   |          |          |
| 18 | 0 | 24 |   |   |          |          |                |             |   |          |          |
| 19 | 0 | 29 |   |   | 0        | 1.96     | MLGetMatrix("m","E19") |         |   |          |          |
| 20 | 4 | 28 |   |   |          |          |                |             |   |          |          |
| 21 | 2 | 17 |   |   |          |          |                |             |   |          |          |
| 22 | 3 | 28 |   |   |          |          |                |             |   |          |          |

Figure 5: Alternative MatLab functions

Figure 6 illustrates an example provided by MatLab with the Excel Link add-in. In this example, the returns over time for three mutual funds is provided in cells B4:D9. In A15, the MLPutMatrix function assigns the column labels from cells F3 through G3 into the array named "Labels."

| | A | B | C | D | E | F | G | H | I | J |
|---|---|---|---|---|---|---|---|---|---|---|
| 1 | Portfolio Efficient Frontier | | | | | | | | | |
| 2 | | | | | | | | Global | Corp. Bnd | Small Cap |
| 3 | Rates of return | Global | Corp. Bnd | Small Cap | | Risk | ROR | Weights | | |
| 4 | Nov-91 | 7.125% | 4.125% | 8.375% | | | | | | |
| 5 | Nov-92 | 5.125% | 5.125% | 3.875% | | | | | | |
| 6 | Nov-93 | -1.375% | 5.750% | 10.500% | | | | | | |
| 7 | Nov-94 | 7.750% | 6.000% | 14.750% | | | | | | |
| 8 | Nov-95 | 8.250% | 6.375% | -3.625% | | | | | | |
| 9 | Nov-96 | 12.625% | 6.125% | 9.125% | | | | | | |
| 10 | | | | | | | | | | |
| 11 | | | | | | | | | | |
| 12 | | | | | | | | | | |
| 13 | Excel Link Functions | | | | | | | | | |
| 14 | 1. Transfer data to MATLAB. | | | | | | | | | |
| 15 | 0 | <== MLPutMatrix("Labels", F3:G3) | | | | | | | | |
| 16 | 0 | <== MLPutMatrix("retseries", B4:D9) | | | | | | | | |
| 17 | | | | | | | | | | |
| 18 | 2. Execute MATLAB Financial Toolbox functions. | | | | | | | | | |
| 19 | 0 | <== MLEvalString("[ret, cov] = ewstats(retseries)") | | | | | | | | |
| 20 | 0 | <== MLEvalString("[risk, ror, weights] = portopt(ret, cov, 20)") | | | | | | | | |
| 21 | | | | | | | | | | |
| 22 | 3. Transfer output data to Excel. | | | | | | | | | |
| 23 | 0 | <== MLGetMatrix("risk", "F4") | | | | | | | | |
| 24 | 0 | <== MLGetMatrix("ror", "G4") | | | | | | | | |
| 25 | 0 | <== MLGetMatrix("weights", "H4") | | | | | | | | |
| 26 | | | | | | | | | | |
| 27 | 4. Plot efficient frontier data and label the figure. | | | | | | | | | |
| 28 | 0 | <== MLEvalString("portopt(ret, cov, 20); grid on; xlabel(Labels{1}); ylabel(Labels{2})") | | | | | | | | |
| 29 | | | | | | | | | | |

Figure 6: MatLab Excel Link example

The next MLPutMatrix function call in cell A16 assigns the data in cells B4 through D9 into the matrix named "retseries". The command in cell A19 calculates the expected return and covariance matrix from the return series.

| | A | B | C | D | E | F | G | H | I | J |
|---|---|---|---|---|---|---|---|---|---|---|
| 1 | Portfolio Efficient Frontier | | | | | | | | | |
| 2 | | | | | | | | Global | Corp. Bnd | Small Cap |
| 3 | Rates of return | Global | Corp. Bnd | Small Cap | | Risk | ROR | Weights | | |
| 4 | Nov-91 | 7.125% | 4.125% | 8.375% | | 0.730% | 5.643% | 0.3% | 96.1% | 3.5% |
| 5 | Nov-92 | 5.125% | 5.125% | 3.875% | | 0.760% | 5.723% | 4.0% | 89.7% | 6.3% |
| 6 | Nov-93 | -1.375% | 5.750% | 10.500% | | 0.844% | 5.803% | 7.7% | 83.3% | 9.0% |
| 7 | Nov-94 | 7.750% | 6.000% | 14.750% | | 0.968% | 5.883% | 11.3% | 76.9% | 11.8% |
| 8 | Nov-95 | 8.250% | 6.375% | -3.625% | | 1.118% | 5.964% | 15.0% | 70.5% | 14.5% |
| 9 | Nov-96 | 12.625% | 6.125% | 9.125% | | 1.287% | 6.044% | 18.7% | 64.0% | 17.3% |
| 10 | | | | | | 1.466% | 6.124% | 22.3% | 57.6% | 20.0% |
| 11 | | | | | | 1.653% | 6.204% | 26.0% | 51.2% | 22.8% |
| 12 | | | | | | 1.846% | 6.284% | 29.7% | 44.8% | 25.5% |
| 13 | Excel Link Functions | | | | | 2.042% | 6.365% | 33.3% | 38.4% | 28.3% |
| 14 | 1. Transfer data to MATLAB. | | | | | 2.241% | 6.445% | 37.0% | 32.0% | 31.1% |
| 15 | 0 | <== MLPutMatrix("Labels", F3:G3) | | | | 2.443% | 6.525% | 40.6% | 25.6% | 33.8% |
| 16 | 0 | <== MLPutMatrix("retseries", B4:D9) | | | | 2.646% | 6.605% | 44.3% | 19.1% | 36.6% |
| 17 | | | | | | 2.850% | 6.685% | 48.0% | 12.7% | 39.3% |
| 18 | 2. Execute MATLAB Financial Toolbox functions. | | | | | 3.055% | 6.766% | 51.6% | 6.3% | 42.1% |
| 19 | 0 | <== MLEvalString("[ret, cov] = ewstats(retseries)") | | | | 3.262% | 6.846% | 55.0% | 0.0% | 45.0% |
| 20 | 0 | <== MLEvalString("[risk, ror, weights] = portopt(ret, cov, 20)") | | | | 3.620% | 6.926% | 41.3% | 0.0% | 58.7% |
| 21 | | | | | | 4.213% | 7.006% | 27.5% | 0.0% | 72.5% |
| 22 | 3. Transfer output data to Excel. | | | | | 4.955% | 7.086% | 13.8% | 0.0% | 86.2% |
| 23 | 0 | <== MLGetMatrix("risk", "F4") | | | | 5.791% | 7.167% | 0.0% | 0.0% | 100.0% |
| 24 | 0 | <== MLGetMatrix("ror", "G4") | | | | | | | | |
| 25 | 0 | <== MLGetMatrix("weights", "H4") | | | | | | | | |
| 26 | | | | | | | | | | |
| 27 | 4. Plot efficient frontier data and label the figure. | | | | | | | | | |
| 28 | 0 | <== MLEvalString("portopt(ret, cov, 20); grid on; xlabel(Labels{1}); ylabel(Labels{2})") | | | | | | | | |
| 29 | | | | | | | | | | |

Figure 7: Portfolio Efficient Frontier Output

The command in cell A20 calculates the mean variance efficient frontier using the vector of expected returns and a specified variance-covariance matrix. The third argument is:

$$[risk, ror, weights] = portopt(ret, cov, 20)$$

The portopt function is the number of portfolios generated along the efficient frontier. The function generates three outputs: portfolio risk ("risk"); portfolio return ("ror"); and the portfolio weights ("weights"). This should be compared to the standard Excel method as described in *Financial Modeling* by Simon Benninga. Benninga's Excel-based method requires eight matrix multiplications, three matrix inversions, and five mathematical calculations that normally fill an entire Excel sheet. These calculations are complex so errors would be easy to make. Add to this, the limitation of only having 52 stocks, which makes Excel not able to perform this calculation in the real world.

The MLGetMatrix commands in cells A23 through A25 write out the portfolio otimization output starting in cells "F4", "G4", and "H4". The final MatLab command in cell A28 plots the output of the potfolio optimization. The graph appears as Figure 8.

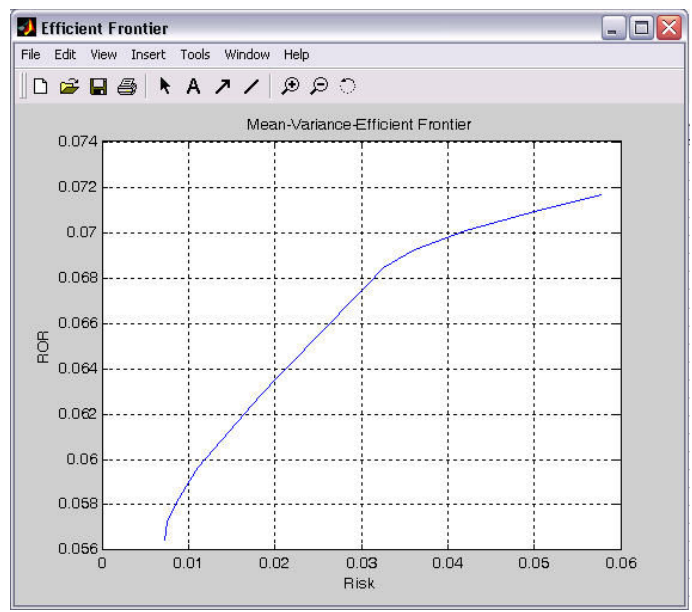

Figure 8: Portfolio Optimization Graphical Output

## VII. Conclusion

Spreadsheets are known to be a great productivity tool for data analysis, model building and reporting for non-programmers. When Excel is used to consolidate financial data, price derivatives or develop financial pro-formas, the use of a proper software design methodology should be employed. There are advantages and disadvantages to using the spreadsheet environment to perform complex calculations. While a spreadsheet is slower than a dedicated software package, a spreadsheet provides the benefit of transparency, allowing for easier verification and validation. But, spreadsheet capability is limited relative to the higher level analytics demanded by users.

For more complex calculations, the robust libraries offered by COTS, such as MatLab, increases spreadsheet capability. Software re-use through COTS components greatly extends the functionality of spreadsheets and at the same time helps achieve better, faster, cheaper spreadsheet development.  But, these components lack the intuitive user development environment, transparency, and furthermore, best practices to control risks.  By creating hybrid applications, EuSpRIG controls flow through to the MatLab code.

While the focus of this paper has been on incorporating MatLab components into a spreadsheet, we believe that EuSpRIG best practices can be applied to to a variety of hybrid spreadsheet/COTS applications to control risks, including those identified by Sarbanes-Oxley and Basel II, arising from complex calculations that were previously beyond the capability of spreadsheets alone.

Futhermore, we believe that EuSpRIG best practices can and should be extended to stand alone MatLab applications as well, since the cause of errors and the error prevention techniques are similar.  The same group of user/programmers are creating complex calculations using higher level tools.